# A Study of Four-Switch Cross-Shaped RIS and A Novel Design Example

Xiaocun Zong,  Binchao Zhang, *Member, IEEE,*  Fan Yang, *Fellow, IEEE,*  Shenheng Xu, *Member, IEEE,* and Maokun Li, *Member, IEEE*

*Abstract*—This paper analyzes the working principle of four-switch cross-shaped reconfigurable intelligent surface (RIS) in detail and reveals the different types of RIS that can be designed based on this structure. Combined with the design examples using this structure in the currently published articles, this paper summarizes and organizes them, and also points out several RIS solutions that have not been designed using this structure. Finally, based on this four-switch cross-shaped structure, this paper proposes a novel RIS design example that can realize the function switching of 1-bit ultra-wideband (UWB) and 2-bit narrowband, and conducts simulation verification. The simulation results show that by optimizing the element structure and controlling the states of the four switches, the 1-bit ultra-wideband function can achieve a frequency band coverage of 10.5GHz-19.8GHz and a 2-bit phase quantization function around 18.12GHz. At the same time, it can realize 60° two-dimensional beam scanning function. We call this novel design "bit reconfigurable metasurface".

## I. INTRODUCTION

RECONFIGURABLE intelligent surface (RIS) is a concept of programmable array consisting of a large number of periodically arranged elements, and is a new kind of phased array. According to the different feed positions, it can be divided into reflect reconfigurable array (RRA) and reconfigurable transmit array (RTA). Thanks to the spatial feeding and integrated phase-tuning technique, RIS has a simple and low-cost structure compared to conventional phased arrays [1]-[3], and thus have become more and more popular in academia and in various civilian and military applications.

The RIS element is adjustable, and the beam scanning function in different directions can be realized by controlling FPGA. In the RIS design, the key to beamforming is to assign the required phase value to each element. The most popular method to realize phase distribution is using PIN diodes [4], [5], the PIN is loaded into the element patch or the phase-shifting structure, and different working modes can be formed by controlling the switch of the PIN. According to the number of phase states, it can be divided into 1-bit, 2-bit and even 3 bit, then we call this phase quatization. Generally, as the number of phase quatization bits increases, the gain and sidelobe level performance improves. However, although the increase in bit number will bring many advantages, it will also bring more obvious element loss, more complex design process and higher production cost. Therefore, how to balance the advantages and disadvantages brought by the increase in bit number is a major issue that needs to be considered during the research process.

In the RIS design, two very important components are the radiation patch and the phase shift structure. The patch is used to receive electromagnetic waves and radiate the phase-shifted electromagnetic signals. There are various radiation patches in the existing literature, such as "U-shaped"[6],[7], "square-shaped"[8],[9], and"circular-shaped"[10]. In theory, any shape can radiate, but in order to better match and radiate bandwidth, some more detailed designs are usually carried out. There are also various solutions for the phase shift structure, such as different resonant states produce different reflection phases[11]-[13], control the length of the microstrip line to use the path difference to produce different phases[14],[15], etc. How to better combine the radiation part and the phase shift part is also a problem that needs to be considered in the design process.

The four-switch cross-shaped patch perfectly solves the problems mentioned above, and it has been used in a large number of RRA/RTA designs due to its simple and symmetrical structure, mature optimization methods, the diverse radiation states and the advantages of integrated "radiation-phase shift" control. This structure includes a metal ground, dielectric substrate layer, a pair of cross-placed metal strips, and extension structures at the end of the metal strips. It also includes four PIN switches. It will have slight deformations depending on different application scenarios. What is even more surprising is that this structure shows different effects when placed parallel to the electric field and when placed at a 45° angle. This feature can also bring different inspirations to our design. When the electric field is incident along the diagonal direction of the element, different

This paragraph of the first footnote will contain the date on which you submitted your paper for review, which is populated by IEEE. It is IEEE style to display support information, including sponsor and financial support acknowledgment, here and not in an acknowledgment section at the end of the article. For example, "This work was supported in part by the U.S. Department of Commerce under Grant 123456." The name of the corresponding author appears after the financial information, e.g. *(Corresponding author: Second B. Author).* Here you may also indicate if authors contributed equally or if there are co-first authors.

The next few paragraphs should contain the authors' current affiliations, including current address and e-mail. For example, First A. Author is with the National Institute of Standards and Technology, Boulder, CO 80305 USA (e-mail: author@ boulder.nist.gov).

Second B. Author Jr. was with Rice University, Houston, TX 77005 USA. He is now with the Department of Physics, Colorado State University, Fort Collins, CO 80523 USA (e-mail: author@lamar.colostate.edu).

Third C. Author is with the Electrical Engineering Department, University of Colorado, Boulder, CO 80309 USA, on leave from the National Research Institute for Metals, Tsukuba 305-0047, Japan (e-mail: author@nrim.go.jp).

resonant states can be formed and various designs can be performed [16]-[25]; When the electric field is incident along the edge of the element [26]-[46], a polarization conversion effect can be formed, and a broadband effect can be formed through various optimizations; combining different states will produce other functions; even combining patches in two directions can produce more unexpected effects. This structure can not only perform phase shift processing on electromagnetic signals, but also can form high-order bit phase quantization numbers based on the original simple structure by controlling the switch state and element angle. If you add some other structures to this structure, you can get even more design inspiration. At the same time, the patch can also radiate, truly achieving an integrated functions of high-order quantization and radiation. Therefore, it is necessary to conduct in-depth exploration and detailed summary of this four-switch cross-shaped patch structure.

This paper will conduct a detailed theoretical analysis of the working principle of this four-switch cross-shaped patch structure, then review and organize different designs, and other possible design solutions are given. Finally, a new design method is proposed based on the theoretical analysis of this paper. The simulation results show that by optimizing the element structure and controlling the states of the four switches, we can realize the function switching of 1-bit UWB and 2-bit narrowband, and the 1-bit UWB function can achieve a frequency band coverage of 10.5GHz-19.8GHz and a 2-bit phase quantization function around 18.12GHz. Our design does not add any other structure to the original 1-bit simple structure, and achieves the function of "bit reconfigurable" under low cost and low design complexity. We have tapped the design potential of the four-switch cross-shaped structure.

## II. BASIC STRUCTURE AND PRINCIPLES

The four-switch cross-shaped RIS has a symmetrical structure and a simple composition. It usually includes a metal ground, a pair of cross-placed metal strips patch, PIN switches on the metal strips, and extension structures at the end of the metal strips, which is generally used to expand bandwidth. If this structure is used in a RTA, the metal ground is replaced by another radiate patch. The relative position relationship between the electric field and the element varies in different application scenarios, as shown in Fig.1.

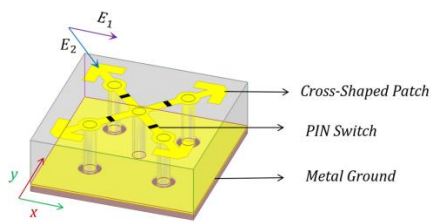

Fig. 1. The general structure of four-switch cross-shaped patch RRA.

*A. The principle of the electric field is incident along the edge of the element*

When the electric field is incident along the edge of the element, just like $E_1$ in Fig1, the patch plays a role in polarization conversion. If the electric field is incident along the $x$ direction, the electric field can be decomposed into two components in the $u$ and $v$ directions. The current components in the two directions produce a 180° phase difference after flowing through the patch, and after synthesis, they become the electric field in the $y$ direction, the principle is shown in Fig.2, this is the specific principle of polarization conversion.

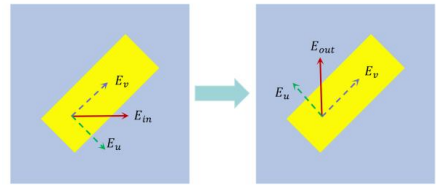

Fig. 2. The principle of polarization conversion by placing patch diagonally

According to the mirror law, when the two patches are placed along ±45°, the electric fields emitted by the two patches will produce an electric field phase difference of 180°. Based on this principle, we can construct a 1-bit RRA element, combining the two patches and controlling the orientation of the patches by loading 4 PIN diodes, as shown in Fig.4: when PIN diodes #1 and #2 are in the OFF state and #3 and #4 are in the ON state, it is state 1; when PIN diodes #1 and #2 are in the ON state and #3 and #4 are in the OFF state, it is state 2.

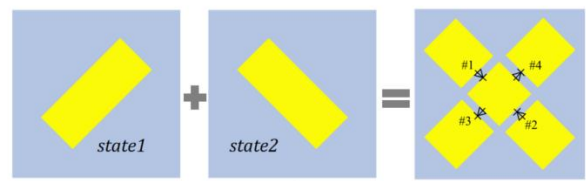

Fig. 3. Schematic diagram of the principle of a 1-bit RRA element

By optimizing the size structure of the element, multi-mode resonance can be used to achieve broadband design. And it is not difficult to find that as long as it is a rectangular strip structure placed along the diagonal and forming an angle of 45° with the electric field, it can form a polarization conversion effect. By using the mirror principle, placing two patches at ±45° will form a phase difference of 180°. It is just that the polarization conversion working frequency produced by different lengths is different. The element under this electric field incidence mode can form three polarization conversion effects at different frequency according to the number of PIN switches opened, that is, the B4, B5, and B6 states in Fig.4.

In addition to the polarization conversion effect, there are still three resonance modes under this incident electric field mode according to the different equivalent resonance lengths. When all PIN switches are in the off state, the resonance state B1 is formed; when the two adjacent branch PIN switches are turned on and the equivalent resonance length is half the side length, it is the resonance state B2; when all PIN switches are turned on, or the two adjacent branch switches are turned on to form an equivalent resonance length of the same length as the side length, this is the resonance state B3, which enriches the design diversity. As shown in Fig.4, all the design possibilities are plotted and the equivalent resonant size can be seen.

| No. | Element Status Diagram | Equivalent Resonant Size |
|---|---|---|
| B1. | | |
| B2. | | |
| B3. | | |
| B4. | | |
| B5. | | |
| B6. | | |

Fig.4. All possible states of the electric field is incident along the edge of the element.

### B. The principle of the electric field is incident along the diagonal of the element

When the electric field is incident along the diagonal of the element, just like $E_2$ in Fig.1, the effect at this time is almost the opposite of the diagonally placed mentioned above. The element mainly works in the resonant mode. Different resonant states can be formed according to the number of PIN diodes turned on. At the same time, this element also has a polarization conversion state, which provides more possibilities for design diversity.

According to the relative position relationship between the cross metal strips and the incident direction of the electric field, different resonance principles can be formed. If the metal strip branch in the switch-on state is perpendicular to the incident electric field, then no matter how complex the structure of the branch perpendicular to the electric field is, they are equivalent to the same resonant state, because the vertical branch does not resonate, and this is state A1. When only one PIN switch is turned on in the branch along the electric field direction, no matter how many PIN switches are turned on in the vertical branch direction, the resonant state at this time is equivalent, which is the resonance principle of A2. Similarly, when two PIN switches are all turned on in the branch along the electric field direction, no matter how many PIN switches are turned on in the vertical branch direction, the resonance state at this time is equivalent, which is the resonance principle of A3.

In addition to the three resonant states mentioned above, when the electric field is incident along the diagonal direction, a polarization conversion effect can also be formed. When two adjacent branch switches are turned on, the current will change the direction of flow, forming a polarization effect. This is the principle of A4. Then we draw all the possible states and corresponding principles in Fig.5.

### III. REVIEW OF DESIGN CASES

From the analysis in section II, we can see that by changing the incident direction of the electric field, the 32 forms of the four-switch cross-shaped element structure can form 10 resonance or polarization conversion modes. Combining these modes, a variety of RA and TA designs can be designed, and elements with different bit numbers can be flexibly designed.

When the electric field is incident along the edge of the element, if only two opposite switches are open (B5), a large number of works have been designed using this principle, that is, the commonly used 1-bit polarization conversion RIS element, references [26],[27],[29]-[34] made a polarization conversion broadband RIS, this principle can be used to design both RA and RA, both of which can achieve ultra-wideband, which is also the most widely used application of this structure, even some design can achieve bandwidth expansion of three times its own center frequency; what's more, the RIS designed using this polarization conversion principle can also be used to generate orbital angular momentum (OAM) beams, references [28] and [35] use this principle to design 1-bit TRA and RRA to generate OAM; there are also a lot of articles using this structure to achieve RCS reduction, [37]-[43] design RIS by exploiting polarization conversion states to suppress the scattering of metal objects by the combination of scattering and reflection, which has significant potential in the applications of antenna designs or stealth technology fields; besides, this design of RIS has also made achievements in the field of communications, reference [44] proposes a broadband metasurface-based wireless communication system that can actively adapt to multiple users located at versatile directions through joint modulation of digital signals in the time domain and wave scatterings in the space domain; moreover, this structure can also be used to design 1-bit transmission array by slightly modifying it, references [19] and [20] designed TA based on this structure to achieve polarization conversion, reference [19] realized a wideband 1-bit filtenna-to-filtenna cross-polarization converter by using multimode resonance, reference [20] used this structure to construct a non-reconfigurable transmission array that can achieve circular-to-linear polarization conversion. Besides, reference [36] adjusted the patch size, and made a pair of dual-frequency reconfigurable polarization conversion reflectarray, which is capable of generating two beams at X- and Ku- bands.

| No. | Element Status Diagram | | | | Equivalent Resonant Size |
|---|---|---|---|---|---|
| A1. | | | | | |
| A2. | | | | | |
| A3. | | | | | |
| A4. | | | | | |

Fig.5. All possible states of placing the element parallel to the E field.

In addition to the commonly used B5 mode, other modes can also produce different effects. In the references [16], a 1-bit independently controllable dual circular polarization RRA was designed by using two resonant states and two polarization conversion states (A2 and A4); reference [18] achieved 1-bit independent control of dual-linear polarization by introducing additional bias, then realizing four states (A1 and A2); in the references [21], this article designs a dual-polarized 2-bit RRA, by introducing an additional PIN diode (A1, A2 and A3); reference [45] uses two polarization conversion states and two resonance states (B1, B3 and B6) to design a 2-bit RRA element by combining these four states; reference [46] uses four states: all PIN diodes are off, two states where only one PIN is on, and one state where two PIN diodes are on (B1, B4 and B5), realized the 1-bit phase resolution in the two polarization direction.

There are also papers that use a combination of two incident directions for design. In reference [24], the authors designed a non-reconfigurable 2-bit RRA, using four states (A1, A3 and B5), which is shown in the Fig.6, and a vector beam modulator is achieved by combining orthogonal polarizations and orbital angular momentum modes. In reference [25], a 2-bit coding metasurface for ultra-wideband and polarization insensitive RCS reduction is designed using the same four states.

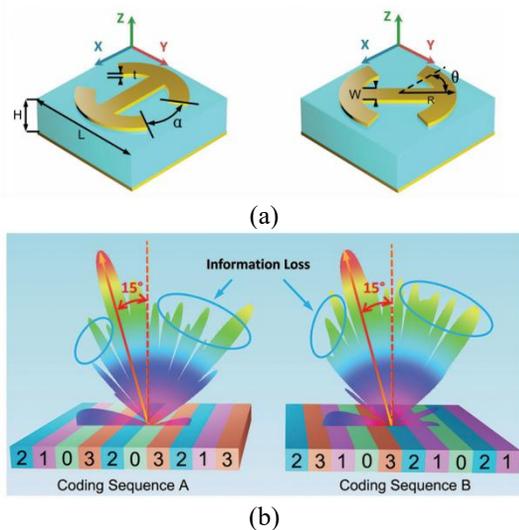

(a)

(b)

Fig.6. The geometry of :(a)coding elements and (b)array in reference[24]

Out of curiosity, what will be the effect of opening another branch on the basis of the polarization conversion effect of B5? That is, state B6. It can be predicted that there must be a certain frequency band that can achieve 100% polarization conversion. At the same time, there will also be a certain frequency band where polarization cancellation occurs due to the opposite directions of the two cross currents, thereby having a certain impact on the reflect amplitude and phase. Polarization cancellation brings loss, but the reflection phase can also be adjusted based on this. We use the two states B5 and B6 to realize a variable bit number reflectarray of 1-bit UWB and 2-bit narrowband, we call this "bit reconfigurable". At the same time, by using the change of patch length design, different polarization conversion states can be combined to design a dual-band ultra-wideband 1-bit element, this is also a novel design idea we proposed. The above two examples are both polarization conversion antennas. According to the principle analyzed above, by introducing an additional PIN switch, we can realize a co-polarization reflectarray with four resonant states.(A1, A2, A3 and another state). We have reviewed the possible designs, as shown in Fig.8.

In short, this four-switch cross-shaped RIS structure can achieve a variety of designs and functions, and there are still many possible designs worth exploring.

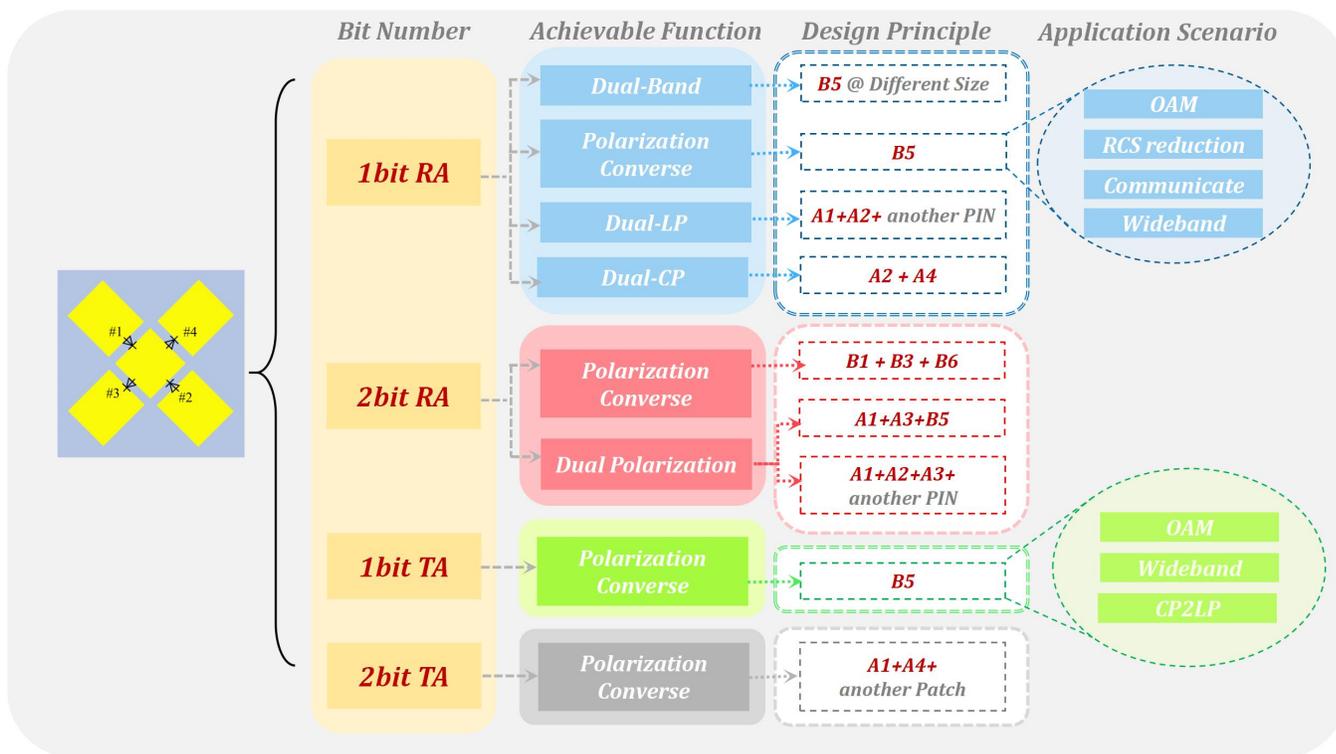

Fig.7. Based on four-switch cross-shaped structure some existing works

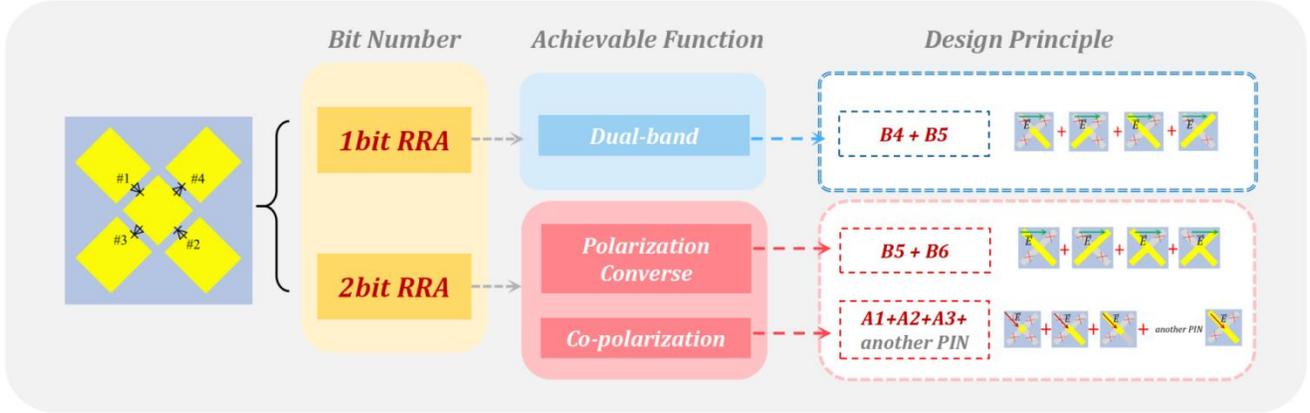

Fig.8. Based on four-switch cross-shaped structure our proposed possible designs

## IV. A NOVEL DESIGN EXAMPLE

Based on the design principle in Section II.A, this section proposes a "bit reconfigurable reflectarray" that can achieve ultra-wideband 1-bit and narrowband 2-bit. As shown in fig.9, the 1-bit principle is achieved by using the mirror principle and polarization conversion effect; on the basis of the original 1-bit, the current cancellation is achieved by introducing the reverse current of the branch, which brings an additional phase effect, thereby generating 2-bit phase quantization.

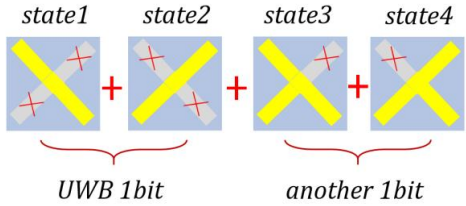

Fig.9. Concept diagram of the four states.

The bit reconfigurable reflectarray element structure includes: a pair of metal strips placed along the diagonal and the extension structure at the end, metal ground, prepreg, dielectric substrate and the bottom DC control circuit. Since the four PIN switches of this design need to be controlled independently, thus four control lines are required, which is also a point different from other designs. The specific details are shown in the Fig.10.

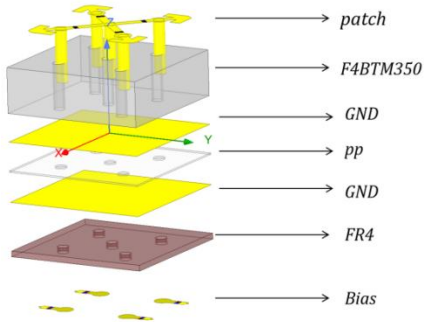

Fig.10. Exploded diagram of bit reconfigurable element structure.

The element is simulated in *HFSS*. When optimizing the element, the 1-bit function is mainly achieved by adjusting the length of the metal strip, the length of the end extension part and the thickness of the upper dielectric substrate, while the 2-bit function is mainly achieved by adjusting the position of the PIN switch to the center point to achieve a 90° phase shift, and some previously set parameters will also be slightly adjusted. From the simulation results by *HFSS*, we can see that the coverage frequency band of 1 bit is very wide, and it can maintain a low loss level from 10.5 GHz to 19 GHz. At the same time, it can achieve a stable 180° phase difference and a bandwidth of 60%.

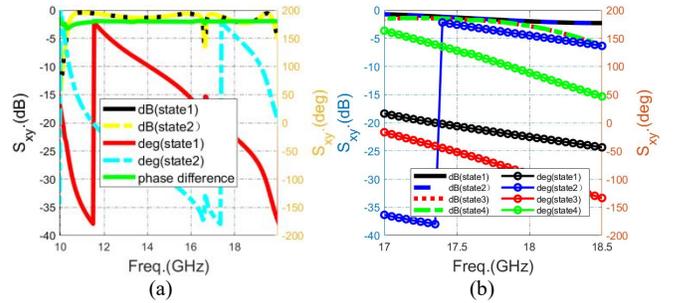

Fig.11. Element simulation results: (a) UWB 1-bit reflect phase and reflection amplitude (b) 2-bit reflect phase and reflection amplitude.

The coverage frequency band of 2-bit is relatively narrow. At around 18.12GHz, the phase difference between state1 and state3 (state2 and state4) can reach about 70°, and the element loss is 2.064dB, 2.067dB, 3.28dB and 3.36dB respectively. The reason for the large element loss at this time is the polarization loss caused by the cancellation of current.

In order to measure the error caused by phase quantization and amplitude loss in element design, the concept of ERA is introduced [47]. The average loss of the element is calculated.

$$ERA = \frac{1}{2\pi} \int_0^{2\pi} \max\{A_i \cos(\phi - \varphi_i)\} d\phi \qquad (3)$$

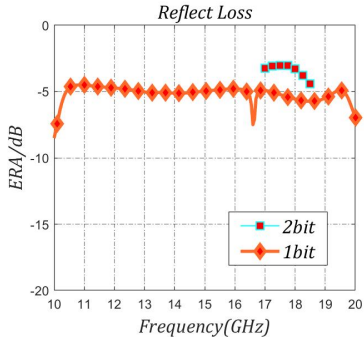

Fig.12. ERA comparison between 1-bit and 2-bit.

Calculations in Fig.12 show that, considering phase quantization loss and element loss, the ERA of 1-bit is controlled within -6dB in the entire frequency band, and the ERA at 18.12GHz is -5.63dB. The ERA of 2-bit can be controlled within -4dB in some frequency bands, and the ERA at 18.12GHz is -3.4dB, which is a 2.23dB performance improvement compared to 1-bit. Therefore, although the element loss of 2-bit is relatively large at this time, the advantage brought by the improvement of phase quantization accuracy is more obvious, and it still has a great advantage over 1-bit.

Since the 1-bit function has been verified in many other papers, only the 2-bit function verification is required. The focal diameter ratio is set to 0.9, and the phase constant has not been optimized. The array size of the full-wave simulation is 16 × 16. The full-wave simulation by *CST* scanned the patterns with a scanning angle of 10° from 0° to 60°, as shown in Fig.17. The maximum aperture efficiency can reach more than 20%, and the side lobe can be controlled below -15dB. Within the 2-bit phase quantization bandwidth of the bit-reconfigurable design, the beam scanning function can be realized. Since the element period exceeds half a wavelength, a serious sidelobe phenomenon occurs during large-angle scanning, which is unavoidable.

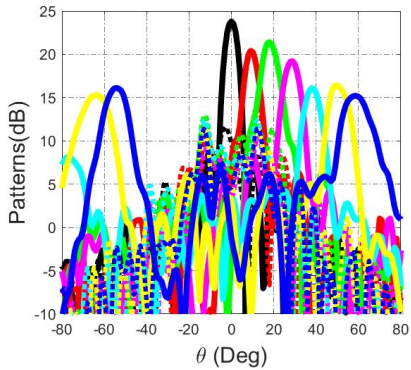

Fig.13. Beam scanning pattern for full-wave simulation

At this time, the focal diameter ratio and phase constant have not been optimized, so the directional pattern at each scanning angle is not the directional pattern with the highest gain. We will continue to optimize the element simulation and array full-wave simulation in the future. The 2-bit aperture efficiency is expected to increase to more than 25%.

At the same time, we will process the prototype for testing. Welcome to pay attention to our subsequent work. Anyway, the concept of "bit reconfigurable reflectarray" has been verified so far.

IV. CONCLUSIONS

This paper systematically reviews the four-switch cross-shaped RIS, including the principle analysis and overview of the two structures of placing the element diagonally and placing the element parallel to the E field, and we proposes several possible design ideas. Finally, based on the principle, this paper proposes the concept of "bit-reconfigurable reflectarray", which can achieve ultra-wideband 1-bit and narrowband 2-bit. The simulation results show that the 1-bit ultra-wideband function can achieve a frequency band coverage of 10.5GHz-19.8GHz and a 2-bit phase quantization function around 18.12GHz. At the same time, the beam scanning function is verified.